\documentclass[a4paper]{lipics-v2021}
\usepackage{mathtools}
\usepackage{cite}
\newcommand{\RE}{\mathbb{R}}

\newcommand{\bdOmega}{\partial \kern+1pt \Omega} 

\newcommand{\SP}{\kern+1pt}             
\newcommand{\revFunk}[1]{{}^r \kern-1pt F_{#1}}

\DeclareMathOperator{\vol}{vol}

\hideLIPIcs
\nolinenumbers

\title{French Onion Soup, Ipelets for Points and Polygons}
\titlerunning{French Onion Soup, Ipelets for Points and Polygons}

\author{Klint Faber}{University of Maryland, College Park, USA \and \url{~}}{kfaber@terpmail.umd.edu }{}{}

\author{Auguste H. Gezalyan}{Department of Computer Science, University of Maryland, College Park, USA \and \url{~}}{octavo@umd.edu}{https://orcid.org/0000-0002-5704-312X}{}

\author{Adam Martinson}{University of Maryland, College Park, USA \and \url{~}}{adammartinson123@gmail.com}{}{}

\author{Aniruddh
 Mutnuru}{University of Maryland, College Park, USA \and \url{~}}{amutnuru@terpmail.umd.edu }{}{}

\author{Nithin Parepally}{Department of Computer Science, University of Maryland, College Park, USA \and \url{~}}{nparepa@terpmail.umd.edu}{}{}

\author{Ryan Parker}{University of Maryland, College Park, USA \and \url{~}}{rpark227@terpmail.umd.edu }{}{}

\author{Mihil Sreenilayam}{University of Maryland, College Park, USA \and \url{~}}{mihilvs@terpmail.umd.edu }{}{}

\author{Aram Zaprosyan}{ University of Maryland, College Park, USA \and \url{~}}{aramzap@umd.edu}{}{}

\author{David M. Mount}{Department of Computer Science, University of Maryland, College Park, USA \and \url{https://www.cs.umd.edu/~mount/}}{mount@umd.edu}{https://orcid.org/0000-0002-3290-8932}{}

\authorrunning{Gezalyan, Chatterjee, Du, Hwang, Mangla, Mount, Parepally, Wu}

\Copyright{Auguste H. Gezalyan, Ainesh Chatterjee, Hongyang Du, Sarah Hwang, Sukrit Mangla, David M. Mount, Nithin Parepally, and Kenny Wu}


\date{\today}

\EventEditors{Erin Chambers and Joachim Gudmundsson}
\EventNoEds{2}
\EventLongTitle{40th International Symposium on Computational Geometry}
\EventShortTitle{SoCG 2024}
\EventAcronym{SoCG}
\EventYear{2023}
\EventDate{June 12--15, 2023}
\EventLocation{Richardson, TX, USA}
\EventLogo{}
\SeriesVolume{39}
\ArticleNo{23}

\begin{document}

\maketitle

\begin{abstract}
There are many structures, both classical and modern, involving point-sets and polygons whose deeper understanding can be facilitated through interactive visualizations. The Ipe extensible drawing editor, developed by Otfried Cheong, is a widely used software system for generating geometric figures. One of its features is the capability to extend its functionality through programs called Ipelets. In this media submission, we showcase a collection of new Ipelets that construct a variety of geometric based structures based on point sets and polygons. These include quad trees, trapezoidal maps, beta skeletons, floating bodies of convex polygons, onion graphs, fractals (Sierpi\'{n}ski triangle and carpet), simple polygon triangulations, and random point sets in simple polygons. All of our Ipelets are programmed in Lua and are freely available.

\end{abstract}

\section{Introduction}
We present several \textbf{Ipelets} for sets of points and polygons in the $\RE^2$ plane. These Ipelets include quadtrees, trapezoidal maps, onion decompositions, random point sampling in polygons, beta skeletons, floating bodies, the Sierpi\'{n}ski carpet and triangle fractals, and simple polygon triangulations. All the Ipelets are programmed in Lua and are freely available at \url{https://github.com/nithin1527/onion-soup}. To install an Ipelet, download the file and place it in the \texttt{ipelets} subfolder of your Ipe folder.

\section{Points}
In this section, we describe the geometric structures that our Ipelets compute on points. There are a few Ipelets available that already build geometric structures on point sets, such as Voronoi diagrams \cite{postechDNN_ipelet, IPE} and Delaunay triangulations \cite{IPE}. We contribute to this with the additions of the following Ipelets.

\subsection{Quadtrees}
\begin{figure}[htbp]
    \centerline{\includegraphics[scale=0.40]{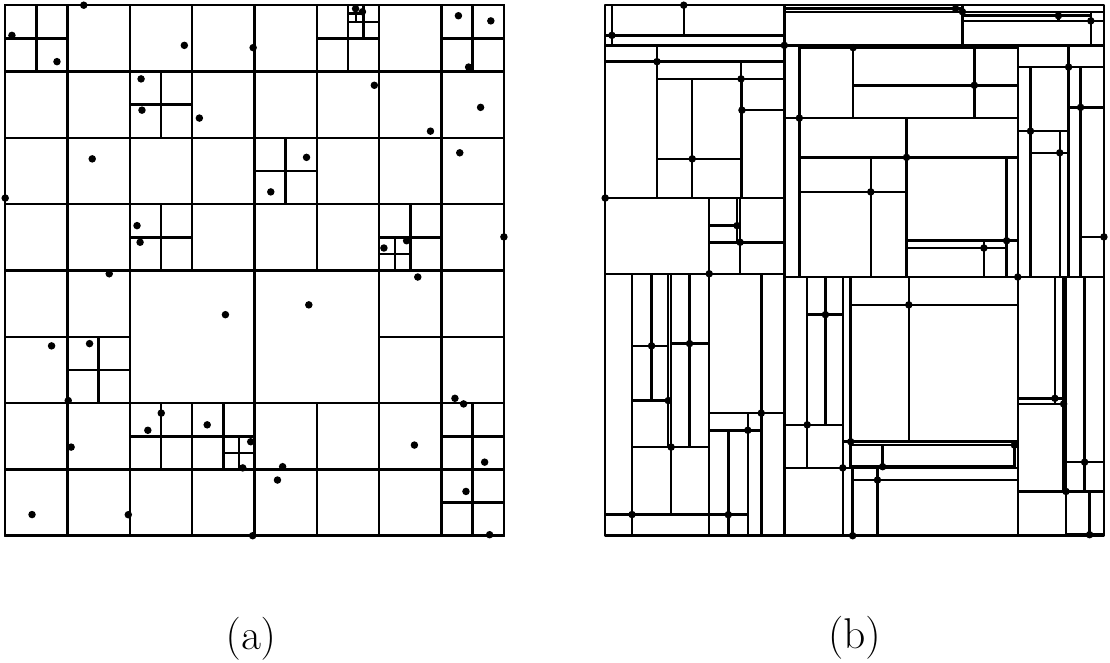}}
    \caption{(a) Point-region quadtree and (b) point quadtree.} \label{fig:Quadtrees}
\end{figure}

A quadtree is a data structure used to continuously divide two-dimensional space into four regions/nodes, usually named northwest, northeast, southwest, and southeast. A point quadtree is a specific type of quadtree in which each node houses a point and its four regions are determined by dividing its region into four sub-regions, with the point itself as the center (see Figure~\ref{fig:Quadtrees}(a)). A point-region quadtree is another type of quadtree in which each node represents a region of space with a certain maximum capacity of points, such that points are only housed in leaf nodes (see Figure~\ref{fig:Quadtrees}(b)). The importance of quadtrees arises from their enabling of efficient operations on data points in two-dimensional space. Consequently, their applications include areas of image representation/processing, mesh generation, and two-dimensional collision detection \cite{deberg2010book,SametStructures}. We contribute an Ipelet for generating both these types of quadtrees, and we produce the array format for its contents in a copyable form.

\subsection{Trapezoidal Maps}

\begin{figure}[htbp]
    \centerline{\includegraphics[scale=0.45]{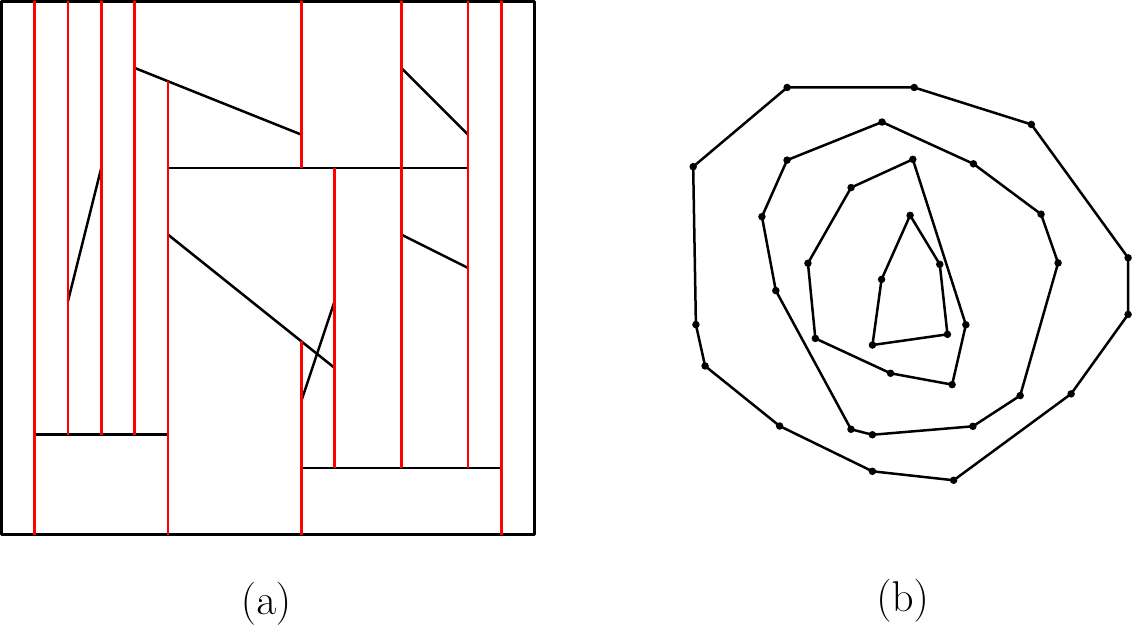}}
    \caption{(a) A trapezoidal map of a set of segments and (b) an onion decomposition of a set of points.} \label{fig:TrapOnion}
\end{figure}

Given a collection of line segments, a trapezoidal map is a subdivision of space that respects these segments. The trapezoidal map is created by shooting vertical line segments vertically up and down from each of the endpoints of the segments until the vertical line intersects another segment or the bounding box (see Figure~\ref{fig:TrapOnion}(a)). In practice, trapezoidal maps allow users to preform planar point location, with respect to the trapezoids that make up the map, in $O(\log(n))$ time. This find applications in many areas, including geographic information systems, computer-aided design, among others \cite{deberg2010book,SametStructures}. We provide an Ipelet which, given a set of line segments and an optional bounding box, compute a trapezoidal map of these line segments. It also provides a method for copying the details of the map in an array format.

\subsection{Onion Decompositions}

\begin{definition}[Onion Decomposition]
Given a finite set of points $P \subset \RE^n$, the onion decomposition of $P$ is the nested sequence of convex sets $\{C_0,C_1,\dots,C_m\}$, where:
\begin{enumerate}
\item[$(1)$] $C_i$ is the convex hull of $P_i$.
\item[$(2)$] $P_0 = P$ and $P_i = P \setminus \bigcup_{j=0}^{i-1} C_j$, for $i \geq 1$.
\item[$(3)$] $P_m$ is the last $P_i$ containing any points.
\end{enumerate}

\end{definition}

This is the sequence of convex polygons that result by repeatedly taking the convex hull of the set, and then removing these points from the set \cite{chazelle1985convex} (see Figure~\ref{fig:TrapOnion}(b)). Onion decompositions of points have found a variety of uses in robust statistics \cite{barnett1976ordering, eddy1982convex} and fractional cascading \cite{chazelle1985power}. We provide software that allows users to select a point set in IPE and generate its onion decomposition.

\subsection{Beta Skeletons}

\begin{definition}[Beta Skeleton]
Given a set of points (vertices) $V$ and a $\beta \in \RE^+$, the beta skeleton of the set of points is the graph $G=(V,E)$ where: There exists an edge $\{v_1,v_2\}\in E$ if and only if for all $v_3 \neq v_1,v_2$, the angle $\angle v_1 v_3 v_2$ exceeds $\theta$, where
\[
    \theta
        ~ = ~ \begin{cases}
                \displaystyle \arcsin \frac{1}{\beta} & \text{if } \beta \geq 1, \\
                \pi - \arcsin (\beta) & \text{if } \beta \leq 1.
            \end{cases}
\]
\end{definition}

\begin{figure}[htbp]
    \centerline{\includegraphics[scale=0.50]{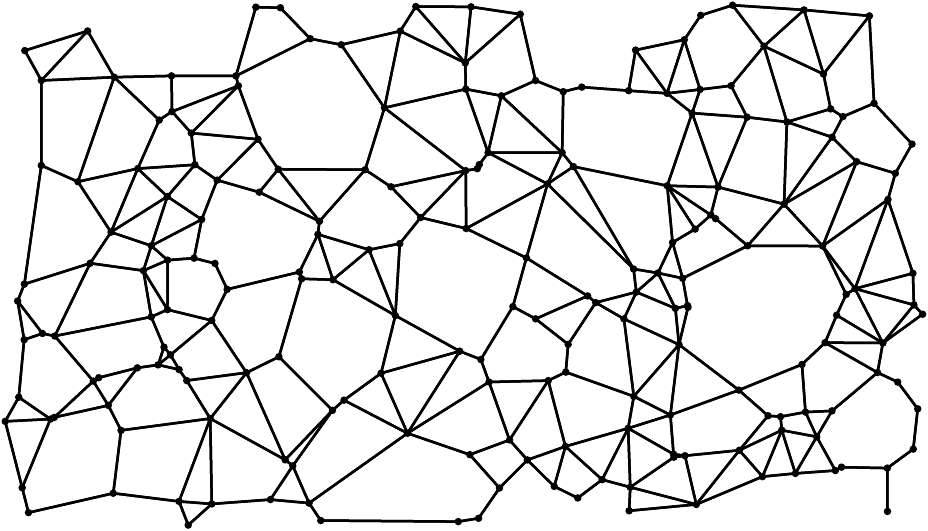}}
    \caption{Beta skeleton for a point set for $\beta=1$, which is equivalent to the Gabriel graph} \label{fig:BetaSkeleton}
\end{figure} 

Note that Gabriel graph and relative neighbor graph (RNG) of the point set arise as special cases when $\beta=1$ and $\beta=2$, respectively. The beta skeleton has applications ranging from machine learning \cite{zhang2002locating} to astronomy \cite{fang2019beta}.

\section{Polygons}
In this section, we describe the geometric structures that our Ipelets compute on polygons, for more, see ``Ipelets for the Convex Polygonal Geometry'' \cite{parepally2024ipelets}. 
\subsection{Floating Body}

Given a convex body $\Omega \subset \RE^d$, a \emph{cap} is the intersection of $\Omega$ with a halfspace $H$. Letting $h$ denote the bounding hyperplane of $H$, the \emph{base} of the cap is $h \cap \Omega$. When $d = 2$, the \emph{area} of a cap is its Lebesgue measure. 

\begin{definition}[Dupin Floating Body]
Given a convex body $\Omega \subset \RE^n$ and $\delta \in [0,1]$, the Dupin floating body, $F_\delta$, is the envelope of the set of points that are the centroids of the bases of all caps that cut off a volume of $\delta \cdot \vol(\Omega)$ \cite{caglar2010floating}.
\end{definition}

\begin{definition}[Convex Floating Body]
Given a convex body $\Omega \subset \RE^n$ and $\delta \in [0,1]$, the convex floating body $K_\delta$ is the
intersection of all halfspaces whose defining hyperplanes cut off a set of volume $\delta \cdot \vol(\Omega)$ from $\Omega$ \cite{schutt1990convex}.
\end{definition}

\begin{figure}[htbp]
    \centerline{\includegraphics[scale=0.50]{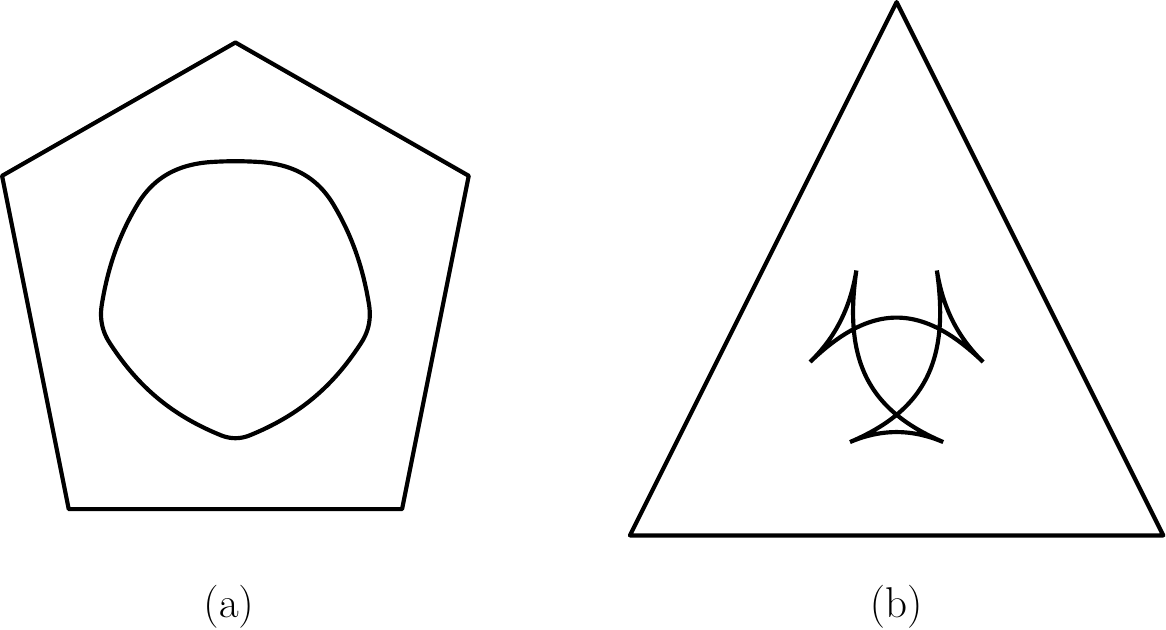}}
    \caption{Two examples of the Dupin floating body, (a) convex and (b) non-convex} \label{fig:FloatingBodies}
\end{figure}

Generally, the Dupin floating body need not be convex (see Figure~\ref{fig:FloatingBodies}(b)), but when it is convex it equals the convex floating body. We provide an Ipelet for generating the Dupin floating body in arbitrary convex polygons (see Figure~\ref{fig:FloatingBodies}). For the purposes of the Ipelet we take $\delta$ as the percentage of the area of the given polygon to cut away with caps. Floating  bodies are used for understanding the boundary structure of convex sets through affine surface area \cite{schutt1990convex}. 

\subsection{Polygon Triangulations and Point Sampling}

\begin{figure}[htbp]
    \centerline{\includegraphics[scale=0.6]{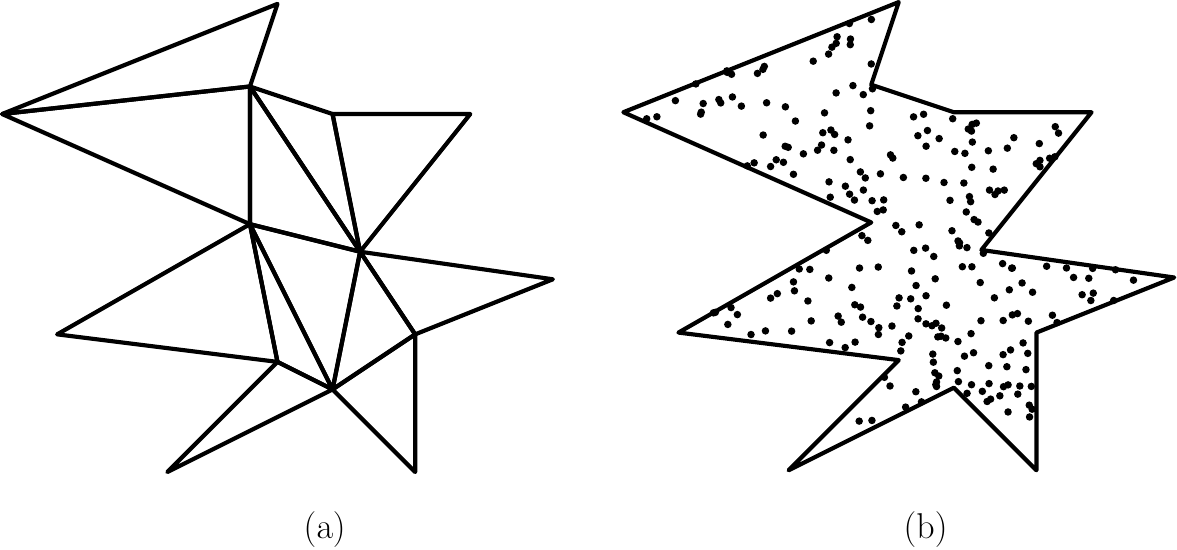}}
    \caption{(a) Triangulating a simple polygon and (b) randomly sampling points in it.} \label{fig:Tri}
\end{figure}
Polygonal triangulations are a ubiquitous tool in computer science \cite{deberg2010book}. Doing this allows efficient point sampling inside non-convex polygons. To do this, points are randomly chosen from inside the triangles (weighted by area) using barycentric coordinates. We provide an Ipelet which, given a polygon, decomposes it into a triangulation (based on code by Nils Olovsson\cite{Olovsson2021}) and randomly places a user specified amount of points in it using this method. 

\subsection{Sierpi\'{n}ski Triangle and Carpet}

The Sierpi\'{n}ski Triangle and Carpet are two self similar fractals defined on simple polygons. They are well known for having a few interesting properties, such as having Lebesgue measure 0 and Haussdorf dimensions $\log_2(3)$ and $\log_3(8)$. Note they contain uncountable many points.

\begin{figure}[htbp]
    \centerline{\includegraphics[scale=0.50]{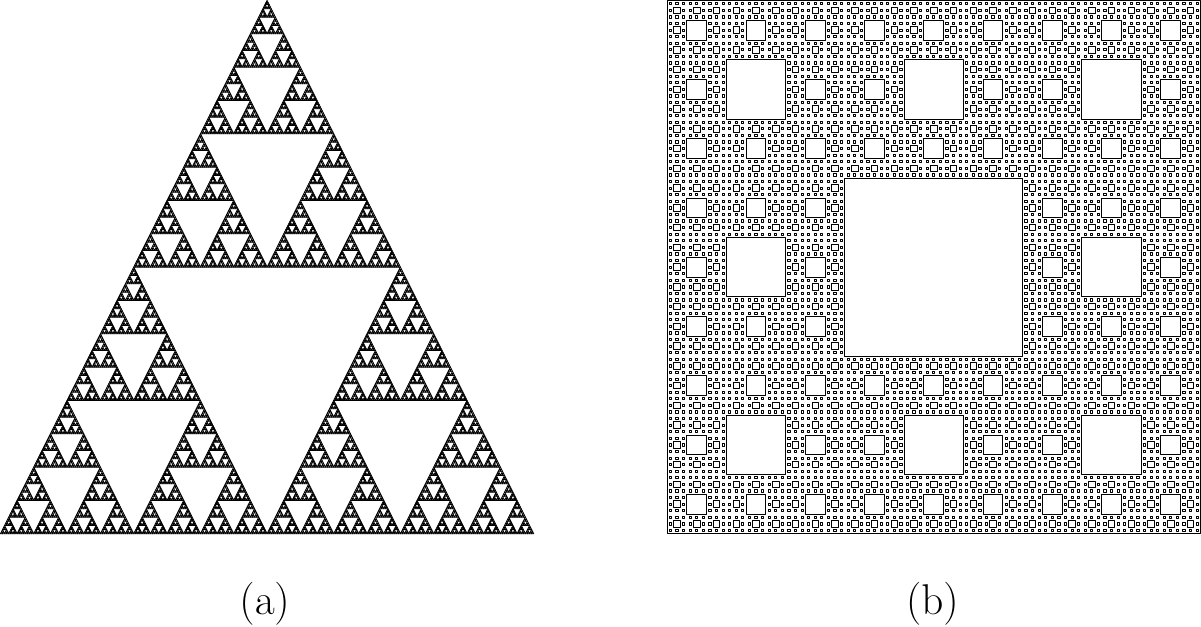}}
    \caption{(a) The Sierpi\'{n}ski Triangle and (b) The Sierpi\'{n}ski Carpet} \label{fig:Sierpinski}
\end{figure}

\bibliography{shortcuts,hilbert}

\begin{thebibliography}{10}

\bibitem{barnett1976ordering}
Vic Barnett.
\newblock The ordering of multivariate data.
\newblock {\em Journal of the Royal Statistical Society: Series A (General)}, 139(3):318--344, 1976.

\bibitem{caglar2010floating}
Umut Caglar.
\newblock Floating bodies.
\newblock Master's thesis, Case Western Reserve University, 2010.

\bibitem{chazelle1985convex}
Bernard Chazelle.
\newblock On the convex layers of a planar set.
\newblock {\em IEEE Transactions on Information Theory}, 31(4):509--517, 1985.

\bibitem{chazelle1985power}
Bernard Chazelle, Leo~J Guibas, and Der-Tsai Lee.
\newblock The power of geometric duality.
\newblock {\em BIT Numerical Mathematics}, 25(1):76--90, 1985.

\bibitem{IPE}
Otfried Cheong.
\newblock The {IPE} extensible drawing editor.
\newblock Version 7.2.28, 2024-03-12.
\newblock URL: \url{https://ipe.otfried.org/}.

\bibitem{deberg2010book}
Mark de~Berg, Otfried Cheong, Marc van Kreveld, and Mark Overmars.
\newblock {\em Computational Geometry: {Algorithms} and Applications}.
\newblock Springer, 3rd edition, 2010.

\bibitem{eddy1982convex}
William~F Eddy.
\newblock Convex hull peeling.
\newblock In {\em COMPSTAT 1982 5th Symposium held at Toulouse 1982: Part I: Proceedings in Computational Statistics}, pages 42--47. Springer, 1982.

\bibitem{fang2019beta}
Feng Fang, Jaime Forero-Romero, Graziano Rossi, Xiao-Dong Li, and Long-Long Feng.
\newblock $\beta$-skeleton analysis of the cosmic web.
\newblock {\em Monthly Notices of the Royal Astronomical Society}, 485(4):5276--5284, 2019.

\bibitem{Olovsson2021}
Nils Olovsson.
\newblock Ear clipping triangulation, 2021.
\newblock Accessed: 2025-02-25.
\newblock URL: \url{https://nils-olovsson.se/articles/ear_clipping_triangulation/}.

\bibitem{parepally2024ipelets}
Nithin Parepally, Ainesh Chatterjee, Auguste~H Gezalyan, Hongyang Du, Sukrit Mangla, Kenny Wu, Sarah Hwang, and David~M Mount.
\newblock Ipelets for the convex polygonal geometry (media exposition).
\newblock In {\em 40th International Symposium on Computational Geometry (SoCG 2024)}, pages 92--1. Schloss Dagstuhl--Leibniz-Zentrum f{\"u}r Informatik, 2024.

\bibitem{postechDNN_ipelet}
postechDNN.
\newblock Ipelet.
\newblock \url{https://github.com/postechDNN/postechDNN/commits/ipelet/dnn/IPELET}, 2022.
\newblock (Commit hash: de82a8d).

\bibitem{SametStructures}
Hanan Samet.
\newblock {\em Foundations of Multidimensional and Metric Data Structures}.
\newblock Morgan Kaufmann Publishers Inc., San Francisco, CA, USA, 2005.

\bibitem{schutt1990convex}
Carsten Sch{\"u}tt and Elisabeth Werner.
\newblock The convex floating body.
\newblock {\em Mathematica Scandinavica}, pages 275--290, 1990.

\bibitem{zhang2002locating}
Wan Zhang and Irwin King.
\newblock Locating support vectors via {/spl beta/}-skeleton technique.
\newblock In {\em Proceedings of the 9th International Conference on Neural Information Processing, 2002. ICONIP'02.}, volume~3, pages 1423--1427. IEEE, 2002.

\end{thebibliography}

\end{document}